\documentclass[%
 reprint,
 amsmath,amssymb,
 aps,
 prl
]{revtex4-2}

\usepackage{graphicx}
\usepackage{dcolumn}
\usepackage{xcolor}
\usepackage{bm}

\begin{document}

\preprint{APS/123-QED}

\title{A fast and streamlined method for the measurement of absolute photodetachment and photodissociation cross-sections}

\author{Salvi Mohandas}
\altaffiliation[Present Address: ]{Institut f\"ur Physik, Johannes Gutenberg-Universit\"at Mainz, Germany.}
\author{Uma Namangalam, Abheek Roy, Hemanth Dinesan}
\altaffiliation[Present Address: ]{Laboratoire de Physique des Lasers CNRS in the Universit\'e Paris 13, France.}
\author{S. Sunil Kumar}%
 \email{sunil.phys@gmail.com}
\affiliation{%
 Department of Physics \& CAMOST, IISER Tirupati $-$ 517619, Andhra Pradesh, India
}%




\date{\today}

\begin{abstract}
The absolute photodetachment cross-section characterizes the photostability of atomic and molecular anions against photodestruction by neutralization. The measurement of this quantity has been reported only for atomic and simple molecular ions. In 2006, Wester’s group introduced a novel ion-trap-based technique to measure the absolute photodetachment cross-section [S. Trippel et al., Phys. Rev. Lett. 97, 193003 (2006)] of $\rm OH^-$. In the present work, we propose a novel methodology to streamline this technique to reduce the measurement time by several orders of magnitude by combining a single experimental rate measurement with a simulated column density distribution of the trapped ions. We validated our approach by reproducing the cross-section reported for $\rm OH^-$ at $632.8\,$nm. Using this technique, we report the first such measurement for a molecule of biological interest, deprotonated indole, at a laser wavelength of 403\,nm. The proposed scheme is anticipated to have a significant and transformative impact on the development of a comprehensive database for photodetachment and photodissociation cross-sections of molecular ions. Furthermore, these measurements have the potential to drive the development of cutting-edge computational codes for cross-section calculations, enabling an unprecedentedly detailed understanding of electron dynamics in large molecules and the light-matter interaction. 
\end{abstract}

\maketitle

\section{\label{sec:level1}Introduction}

Photodetachment, the removal of an electron from a negative ion, is one of the most fundamental processes of the interaction of light with matter. Investigation of this process allows us to examine subtle electron-electron correlations that are responsible for the stability of the anion~\cite{Blondel1996-pk,Chen2008-yt,Compagnon2010-cu}. The \emph{absolute} photodetachment cross-section provides an absolute measure to compare the photostability among various anionic species, even an atomic ion versus a complex molecular ion. The photostability of molecular ions of biological relevance, such as nucleotides and amino acids, has implications for the formation of the first molecules of life on Earth because these molecules would have been exposed to intense ultraviolet light due to the absence of a dense protective atmosphere like the one we have at present~\cite{Cnossen2007-tu}. It also has implications in astrophysics, characterizing the abundance of various molecular species in the interstellar medium, planetary atmospheres, and other astrophysical environments~\cite{Best2011-gw,Kumar2013-ar}. Furthermore, the absolute photodetachment cross-section may serve as a parameter for characterizing the chirality of complex molecular species when its measurement is carried out with circularly polarized light of opposite helicity~\cite{Kruger2021-fy}.

Large deprotonated molecular ions, such as those resulting from an electrospray ionization (ESI) source, are negative ions that belong to a different class of anions because their overall electronic configuration remains the same as that of the neutral species. The photodetachment process from such molecular ions has been investigated by several research groups~\cite{Nelson2018-le,Anstoter2016-gr,Bochenkova2017-rs,Noble2020-sk,Kruger2021-fy,Forbes2011-gu,Forbes2009-aj,Henley2019-qk,Mooney2012-cj,Bravaya2013-lg}. These investigations have explored photoelectron spectra, photoelectron angular distributions, various deactivation mechanisms, and the effect of structural modifications on the photodetachment process. Combined with complementary theoretical investigations, these measurements aid in providing a deeper insight into the intricacies of electron correlation effects. However, a quantitative way of assessing the photostability of these molecular ions in terms of the absolute photodetachment cross-section has not been addressed in the literature.

Measurement of absolute photodetachment cross-section using conventional techniques is a notoriously difficult task, because of which there exist only very few direct measurements~\cite{Smith1955-oa,Branscomb1955-av,Branscomb1958-tb,Champeau1998-jt,Hodges1981-em,Genevriez2018-bq,UnknownUnknown-re}. The majority of experiments measured relative cross-sections, while those reporting absolute cross-sections employed intricate normalization procedures, which were dependent on prior absolute measurements~\cite{Smith1955-oa,Branscomb1955-av,Branscomb1958-tb}. The latter measurements are susceptible to large errors because of the challenges involved in the calibration procedures. A novel technique for measuring the absolute photodetachment cross-section of molecular ions directly, without employing any normalization procedure, was outlined by Wester's group to measure the cross-section of ${\rm OH^-}$~\cite{Trippel2006-ez}. A detailed account of the measurement procedure is discussed in subsequent publications~\cite{Hlavenka2009-cc,Best2011-gw, Kumar2013-ar}. Therefore, only a brief overview is presented here. The $\rm OH^-$ ions were confined in a 22-pole radiofrequency ion trap for specific storage durations before extracting them onto a detector to measure the ion signal intensity. As the storage duration increased, the extracted ion signal intensity decreased exponentially due to various ion loss processes~\cite{Wester2009-mk}, and the background decay rate ($k_{bg}$) was measured. When a focused laser of wavelength $632.8\,$nm was sent through the ion cloud after the ions were loaded into the trap, the ion lifetime decreased owing to the photodetachment of $\rm OH^-$ ions, yielding the corresponding decay rate $k_{tot}$. The photodetachment decay rate was then calculated as $k_{pd}=k_{tot}-k_{bg}$. This measurement procedure was repeated multiple times to enhance statistical reliability.

To measure the absolute photodetachment cross-section, this method involves generating a map of the photodetachment decay rate ($k_{pd}$) of the ions as a function of the position of the interaction of the laser beam with the ion cloud. The ratemap so measured in the publication~\cite{Hlavenka2009-cc} had a grid size of $28\times28$, requiring rate measurements for more than 700 points. This rate map is then integrated and divided by the photon flux ($\Phi$) to obtain the absolute photodetachment cross-section:
\begin{eqnarray}
    \sigma_{pd}=\frac{\int\int dz dy k_{pd}(z,y)}{\Phi},
\end{eqnarray}
where it is assumed that the laser is scanned over the ion cloud in the $zy-$plane. This technique has been employed to obtain the absolute cross-sections of $\rm O^-$~\cite{Hlavenka2009-cc}, carbon chain anions such as $\rm C_nH^- (2,4,6)$ and $\rm C_nN^- (n=1,3)$~\cite{Best2011-gw,Kumar2013-ar}. In all of these experiments, the technique required measuring the decay rate of the ions with a laser pointing at various positions that covered the entire ion cloud. In many of these measurements, obtaining the rate map required sampling of nearly 1000 points. Such an experiment can be very time-consuming, especially when the photodetachment cross-section is very low.

In this work, we propose a methodology that requires measuring only a single-point decay rate and then using numerical simulations to obtain the rate map, thereby reducing the time for such a measurement by several orders of magnitude. We employ this methodology to report the first measurement of the absolute photodetachment cross-section of a molecular ion of biological relevance, deprotonated indole (d-indole), at a laser wavelength of 403\,nm. We \textit{emphasize} that the same technique can be used to measure the absolute \emph{photodissociation} cross sections of any molecular ion (both positive and negative) by counting only the parent ions extracted from the ion trap.

\section{Experimental details}
We carried out these measurements using a home-built experimental setup that features an ESI source integrated with a 16-pole ion trap setup, which has been described in detail elsewhere~\cite{Salvi2023-io}. Briefly, the deprotonated molecular ions of indole are produced in the ESI source and transported through a differentially pumped vacuum system consisting of an ion funnel, a quadrupole ion guide, and a quadrupole mass filter before loading the ions of our interest into a 16-pole ion trap. The ions are buffer-gas-cooled to room temperature using helium. The background lifetime of the ions was found to be around 60\,s, a relatively long lifetime of trapped ions at room temperature, which allows us to measure even the slow decay of ions within the ion trap due to laser photodetachment.

\begin{figure}[h]
\includegraphics[scale=0.30]{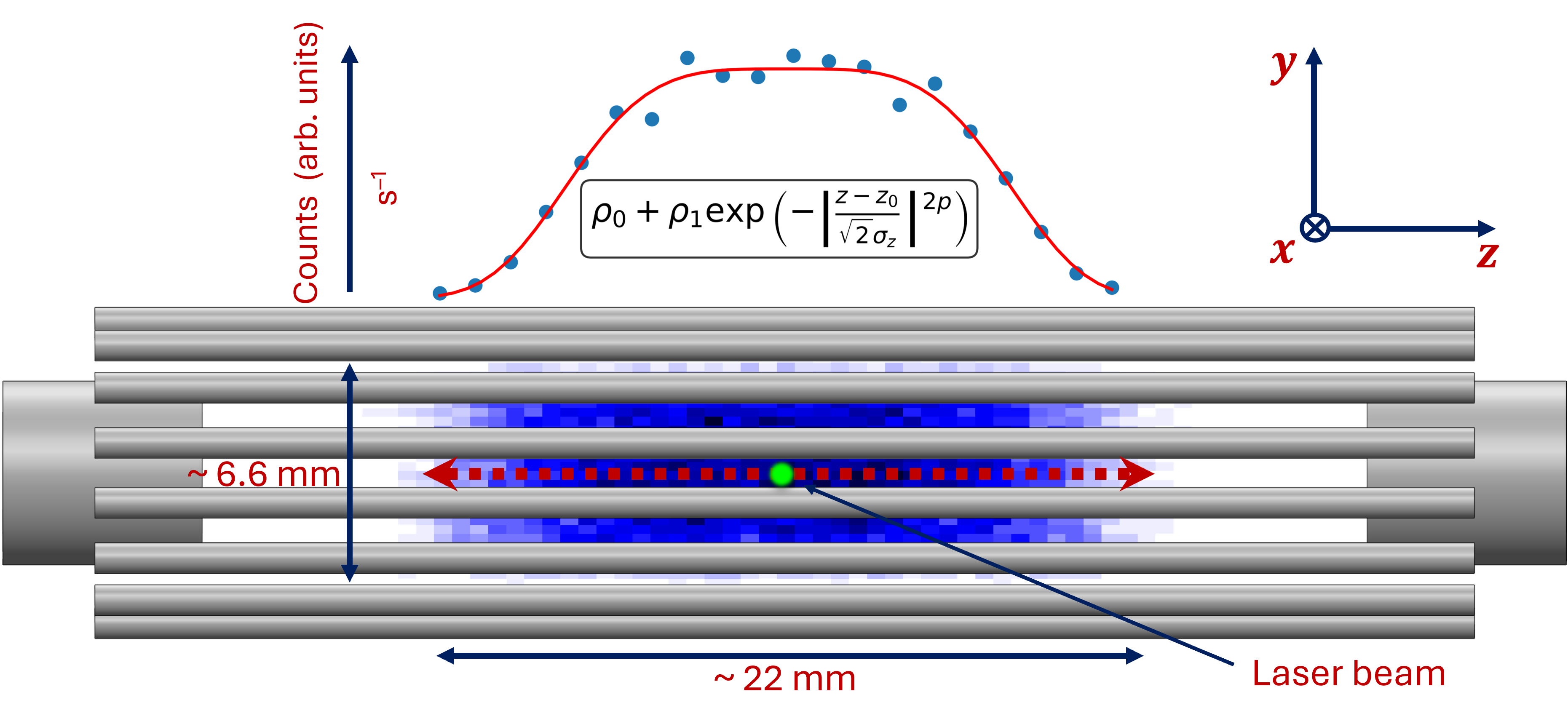}
\caption{\label{fig:scheme_pdxsec} Schematic of the ion trap showing the arrangement for the measurement of the photodetachment decay rate of ions. The 16-pole trap and the ions' distribution (blue histogram) are to scale. The photodetachment decay rate is measured by directing a narrowly-focused laser beam along the $x-$axis so that it interacts with the ion cloud between the two central rods of the trap. The decay rate as a function of the position of the interaction of the laser with the ion cloud is obtained by scanning the laser position along the $z-$axis. The simulated ion column density distribution at $y=0$ ($zx-$plane at the middle of the trap), as shown above the ion trap, follows a flattop Gaussian distribution.}
\end{figure}

\section{Numerical simulations using SIMION}
In our experimental scheme, the ions are optically accessed in a direction transverse to the axis of the ion trap (Fig.~\ref{fig:scheme_pdxsec}). The ion trap is made of 16 stainless steel rods, each with a diameter $1\,$mm, placed to form a cylindrical cage of an inscribed diameter of $9\,$mm. Since the rods obstruct a significant portion of the ion cloud, measuring the complete rate map in the transverse direction is impossible. This is indeed the case for the conventional 22-pole ion trap. We note that the rate measured with the laser interacting with the ions at a given position, say ($z_L,y_L$), on the ion cloud is proportional to the fractional ion column density at that position~\cite{Trippel2006-ez}:
\begin{eqnarray}
    \rho(z_L, y_L)=\frac{n(z_L,y_L)}{N}=\frac{k_{pd}(z_L,y_L)}{\iint dzdyk_{pd}(z,y)},
\end{eqnarray}
where $n(z_L,y_L)$ is the ion column density at $(z_L,y_L)$, $N$ is the total number of ions, and $k_{pd}(z_L,y_L)$ is the rate of decay of the ions from the trap when the laser irradiates the ion cloud at $(z_L,y_L)$. The ion column density is molded by the effective potential the ions experience within the ion trap. Therefore, the rate map can be generated by numerically simulating the ion column density within the trap once the ions are thermalized. To achieve this, we perform ion trajectory simulations inside the ion trap using the SIMION software package with the same electric field characteristics employed in the experiment~\cite{noauthor_undated-vd}. In our simulations, an ion of deprotonated indole is initialized from the trap's center with a Gaussian energy distribution of 0.1\,eV with a standard deviation of 0.02\,eV and allowed to emerge through a half angle of $10^\circ$ with reference to the axis of the trap. The helium buffer gas density inside the ion trap is defined to be $\sim 10^{12}\,{\rm cm}^{-3}$ to match our experimental conditions. It may be noticed that the initial parameters chosen for the ion do not affect the final ion column density distribution, because the thermalizing collisions with the buffer gas erase any information about the initialization. The collisions between the ion and the gas molecules are modeled via a hard-sphere collision model, HS1~\cite{noauthor_undated-vd} at 293\,K. The simulations are carried out with an RF amplitude of $35\,\rm Vpp$ and a frequency of $1.5\,\rm MHz$. The ion trap was floated at $-2.9\,\rm V$. To accurately replicate the experimental ion cloud distribution, it is essential to match all relevant parameters with those used in the experiment. The critical factors include the precise geometric dimensions of the ion trap, the applied RF and DC potentials, the buffer gas temperature, and a sufficiently long storage duration (verified through numerical simulations to ensure thermalization$-$10\,ms in the present case). Additionally, a statistically significant number of ions must be simulated (over 20,000 in the current study) to achieve reliable results. 

Our previous studies have shown that the ions in the mass range relevant to our experiment get thermalized with the buffer gas within 10\,ms after ions are loaded into the trap~\cite{Rajeevan2021-th}. Therefore, we recorded the position of the ion at 15\,ms, well beyond the duration for thermalization. The procedure is repeated by simulating the trajectories of at least twenty thousand ions. The resulting three-dimensional position data is used to generate a two-dimensional histogram representing the ion column density distribution in the $zy$-plane. This distribution, shown in Fig.~\ref{fig:scheme_pdxsec}, is presented alongside a CAD rendering of the ion trap, with the ion cloud displayed beneath the ion trap design. Both horizontal ($z-$axis) and vertical ($y-$axis) slices of this distribution are found to be well-represented by a flattop-Gaussian-type function of the form:
\begin{eqnarray}
    \rho(j)=\rho_0+\rho_1\exp\left[-\left(\frac{|j-j_0|}{\sqrt2\sigma_j}\right)^{2p}\right],\label{eq:ftg1d}
\end{eqnarray}
where $j$ stands for either $z$ or $y$, $j_0$ is the center of the distribution, $\sigma_j$, a measure of the width, and $p$, the flatness of the distribution. Although the simulated distribution can be directly converted into a rate map, relying on a single central point from a distribution with finite statistics may introduce bias. To avoid this, we fit the simulated 2D histogram using a flattop Gaussian function of the form:
\begin{eqnarray}
\begin{split}
\rho(z,y)=\rho_0+\rho_1\exp\left[-\left(\frac{|z-z_0|}{\sqrt2\sigma_z}\right)^{2p}-\left(\frac{|y-y_0|}{\sqrt2\sigma_y}\right)^{2q}\right].\label{eq:ftg2d}   
\end{split}
\end{eqnarray}
Here, the values of $\rho_0$, $\rho_1$, $z_0$, $y_0$, $p$, $q$, $\sigma_z$, and $\sigma_y$, are the fitting parameters. This fitted density distribution is then scaled using the rate we measured at the center of the trap so that the ion column density distribution is transformed into a rate map. We claim that the rate map so constructed should resemble the one that is measured by having the laser beam pass through various positions on the ion cloud. The claim is substantiated by the dependence of the fractional ion column density and the fractional decay rate measured at the same position as illustrated in eq.~\ref{eq:ftg1d}. The volume under the rate map is then computed, which is equivalent to $\int\int dz dy k_{pd}(z,y)$. This value is then divided by the photon flux used in the experiment to compute the absolute photodetachment cross-section. This procedure yielded the photodetachment cross-section of $(1.9\pm0.2)\times10^{-18}\,{\rm cm^2}$ for deprotonated indole at 403\,nm. Notice that a single rate measurement for d-indole takes about an hour. Had we followed the conventional method of measuring the ratemap, it would take about 100 days of continuous measurements under ideal experimental conditions. Thus, our methodology accelerates the measurement process by approximately three orders of magnitude.

\section{Results and discussion}
\begin{figure}[h!]
\centering
\includegraphics[scale=0.28]{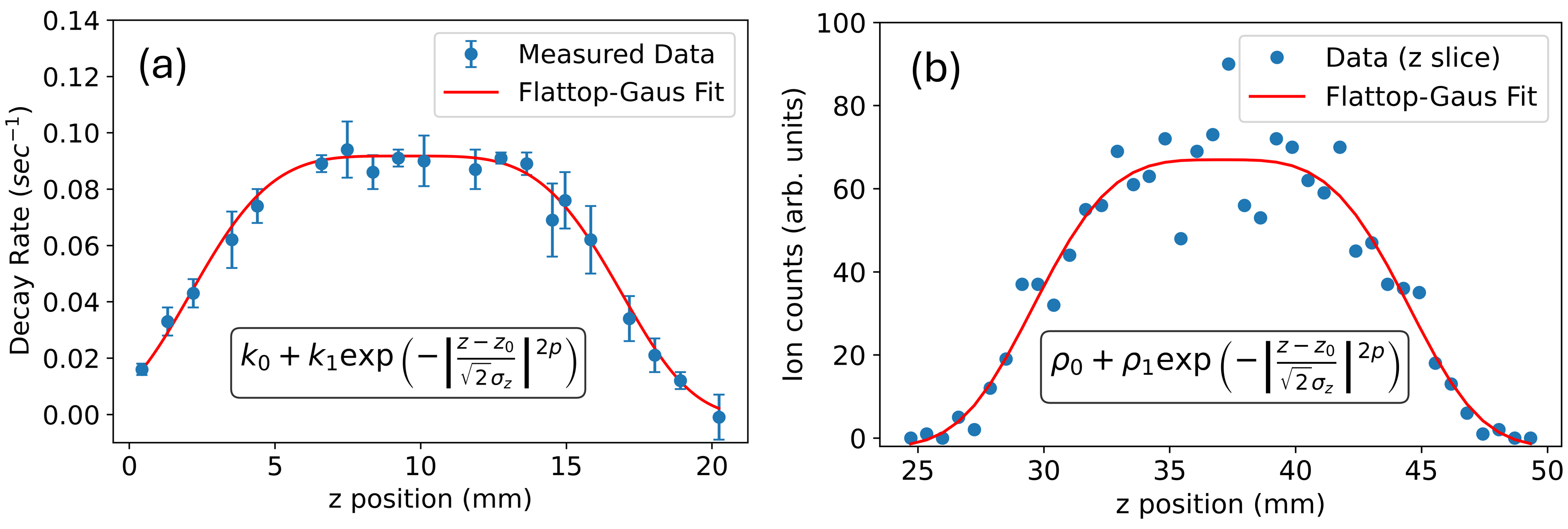}
\caption{\label{fig:pdrate_indole} (a) The measured photodetachment decay rate of d-indole obtained by scanning the laser position along the $z-$axis (the laser beam is directed along the $x-$axis). The distribution is fitted by a flattop Gaussian function given by Eq.~\ref{eq:ftg1d} with $k$ replacing $\rho$. (b) The simulated ion density distribution within the trap along the $z-$axis. The distribution is fitted with the same flattop Gaussian function.} 
\end{figure}

To validate our claim that the numerically simulated ion column density distribution reproduces the measured decay rate map, we measured the decay rate of d-indole by directing the laser through the ion cloud at various $z-$positions as indicated by the double-headed red-colored arrow in Fig. \ref{fig:scheme_pdxsec}. The result of this measurement is depicted in Fig.~\ref{fig:pdrate_indole}(a). The rate curve is fitted with the distribution function given by Eq.~\ref{eq:ftg1d}, where $k$ replaces $\rho$. Fig.~\ref{fig:pdrate_indole}(b) depicts the equivalent plot from the simulation fitted with the same distribution function. Note that the value of $\sigma$, which is a measure of the width of the distribution, agrees well between experiment ($5.6\pm0.3$\,mm) and simulation ($5.8\pm0.3$\,mm) within the error bars. A slice of the simulated distribution along the $y-$direction at $z=0$ also featured a flattop Gaussian function. In this case, the width of the distribution turned out to be much smaller ($\sigma_y=1.9\pm0.1\,\rm mm$), as one would expect.

\begin{figure}[h!]
\centering
\includegraphics[scale=0.27]{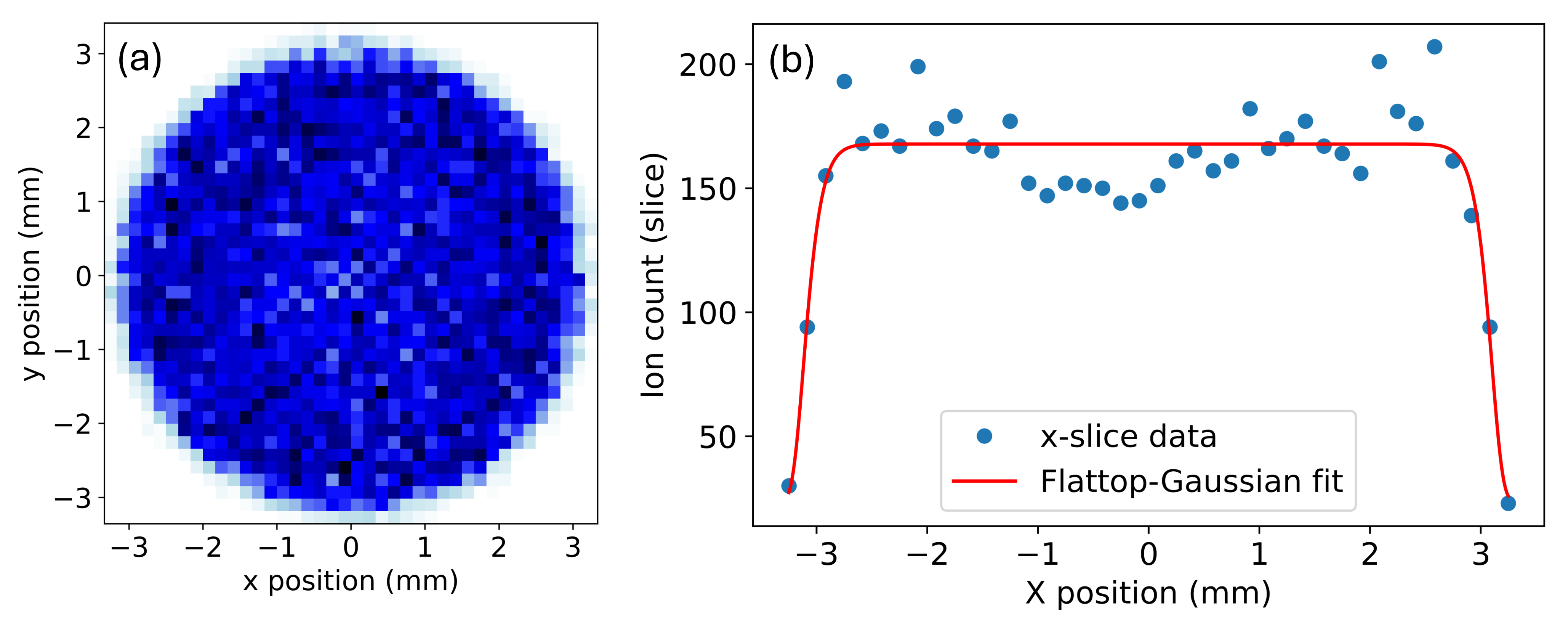}
\caption{\label{fig:iondist-oh} (a) The simulated ion density distribution (projected along the axis of the ion trap) for $\rm OH^-$ ions within a 22-pole ion trap operated at 5\,MHz at an RF amplitude of 150\,V. (b) A slice of the distribution along the $x-$axis, fitted by a flattop Gaussian function given by Eq.~\ref{eq:ftg2dr}, where the value of $m$ that fits best was found to be equal to 16.} 
\end{figure}

To further establish the robustness of our methodology, we employed our technique to reproduce the photodetachment cross-section of $\rm OH^-$ based on the measured rate reported in Trippel et al.~\cite{Trippel2006-ez}. In this experiment, which was performed using a 22-pole ion trap at a temperature of 170\,K, the ion cloud was accessed by directing the laser along the trap axis ($z-$direction). Therefore, the ion cloud distribution is expected to have a circular symmetry due to which they measured the rate just along a diameter. The measured rate featured an unexpected asymmetric bimodal distribution. The bimodal distribution was attributed to the effect of static potential applied to the endcaps and the asymmetry to the field inhomogeneities due to potentials applied external to the trap. In fact, later it was realized that such distributions could occur due to small imperfections while constructing the trap~\cite{Otto2009-ji}. Despite the asymmetric, bimodal distribution, an average value of the rate was used for computing the cross-section in Trippel et al.~\cite{Trippel2006-ez}. In Fig. 3 of this reference, the dashed line represents the expected rate distribution with no endcap effects and imperfections. We used this rate value at the center of the distribution ($\rm 0.225\,s^{-1}$) for constructing the rate map from the simulated ion column density distribution. Since the frequency and amplitude of the radiofrequency (RF) signal and the endcap voltage used for the ion trap were not mentioned in this paper, we used an RF frequency of 5\,MHz at an RF amplitude of 150\,V, with an endcap potential of $-5\,$V as reported in a subsequent paper from the same group~\cite{Hlavenka2009-cc} which included a re-measurement of the cross-section for $\rm OH^-$. The simulated distribution, shown in Fig.~\ref{fig:iondist-oh} is similar to the one that Trippel et al. measured in the original publication~\cite{Trippel2006-ez}, apart from the noted asymmetries and irregularities. The 2D histogram of the ion column density distribution was then fitted with a flattop Gaussian function of the form
\begin{eqnarray}
    k(x,y)=k_0+k_1 \exp\left[-\left(\frac{(x-x_0)^2+(y-y_0)^2}{2\sigma_r^2}\right)^{p}\right],\label{eq:ftg2dr}
\end{eqnarray}
where the value of $\sigma_r$ yields a measure of the size of the ion cloud distribution in the radial direction. We obtained the ion density distribution with these parameters, and the computed cross-section at 632.8\,nm is $(7.3\pm0.7)\times10^{-18}$\,$\rm cm^2$, which is consistent with the value ($(5.6\pm1.4)\times10^{-18}\,\rm cm^2$) reported by Wester's group~\cite{Trippel2006-ez}. It is worth noticing that in a later work from the same group~\cite{Hlavenka2009-cc}, where the entire ion cloud was sampled, the cross-section was revised to be euqal to $8.1(1)_{\rm stat}(7)_{\rm sys}\times10^{-18}\,\rm cm^2$, consistent with our results.

While the technique we introduce here is simple and straightforward to implement, there are a few caveats. In previous works,~\cite{Otto2009-ji,Hlavenka2009-cc} it has been found that at low temperatures, the spatial distribution of the ion cloud was severely distorted due to even minor imperfections in the construction of the trap. In this case, since the ion cloud distribution cannot be accurately simulated due to the exact nature of the imperfection, the cross-section obtained is likely to be prone to large errors. These distortions can worsen if high RF amplitudes and large endcap potentials are used to store the ions. Therefore, it is advisable to operate the ion trap at ``mild" operating conditions (low RF amplitude, endcap voltages, etc.), preferably at room temperature, where the ion cloud distortions are minimal so that an accurate value of the cross-section can be determined.

We find that the method used to determine photodetachment cross-sections is equally applicable for measuring absolute photodissociation cross-sections of molecular ions, regardless of their charge state. In an ion trap, photodissociation typically produces one ionic fragment with a lower $m/z$ and one or more neutral fragments, which escape the trap. If the extraction region is equipped with a high-resolution mass spectrometer (e.g., quadrupole or time-of-flight), a mass spectrum can be recorded by extracting the trapped ions. This allows distinction between parent and fragment ions, provided the trap parameters and fragment energies permit co-trapping of the ions. By monitoring the decay of the parent ion signal over time, the absolute photodissociation cross-section can be determined using the same methodology as for photodetachment. Additionally, if fragment ions are co-trapped and resolved, the growth rate of their signal can be measured, enabling the determination of absolute cross-sections for individual fragmentation channels. These cross-sections reflect the formation rates of specific fragment ions and are particularly valuable for modeling the formation and destruction pathways of molecular ions in the interstellar medium.

\section{Conclusion}

In conclusion, we propose a hybrid methodology (combining experimental data with a numerically simulated ion cloud distribution) for measuring the absolute photodetachment cross-section of any atomic/molecular anion that can be stored inside a multipole RF ion trap. The proposed method reduces the measurement time by several orders of magnitude (ignoring the time required for the simulations, which typically takes a week) compared to that of the existing techniques. If the decay rate due to photodetachment is fast, the cross-section can be measured within a few minutes, barring the amount of time required for arranging the ion generation, trapping, setting up the laser, etc. Therefore, this technique can be employed to perform absolute cross-section measurements as a routine procedure. Furthermore, this method is applicable for measuring absolute \emph{photodissociation} cross-sections if the parent ions remaining within the trap can be distinguished from other photofragmentation products during extraction using a quadrupole mass filter or a time-of-flight spectrometer with a reasonable resolution. In principle, even channel-wise photodissociation cross-sections can be measured using the same methodology if a high-resolution mass spectrometer can be used to measure the rate of production of fragment ions. In short, the proposed scheme can have an unprecedented impact on generating a database of photodestruction cross-sections of atomic and molecular ions of any charge state. Such databases are invaluable for modeling the molecular abundances in astrophysical environments and for assessing the resilience of molecules of life against photodestruction. Since theoretical treatments for computing absolute photodetachment cross-sections are limited to only smaller molecular systems, the new measurements have the potential to stimulate the development of more sophisticated methods that can treat light-molecule interactions involving complex molecular ions. Such advancements would, in turn, provide deeper insights into the electron dynamics within complex molecular species and their interaction with light. 

\section{Acknowledgments}
\begin{acknowledgments}
Funding: SERB Core Research Grant (SERB CRG) $-$ CRG/2021/005022. We gratefully acknowledge the critical feedback given by Professor E. Krishnakumar, Raman Research Institute Bengaluru, on this work.
\end{acknowledgments}

\bibliography{pdxsc_indole_arxiv}

\end{document}